# Chromium on Mercury: New results from the MESSENGER X-Ray Spectrometer and implications for the innermost planet's geochemical evolution

Larry R. Nittler[1,2*], Asmaa Boujibar[1,3], Ellen Crapster-Pregont[4,5], Elizabeth A. Frank[1], Timothy J. McCoy[6], Francis M. McCubbin[7], Richard D. Starr[8,9], Audrey Vorburger[4,10], Shoshana Z. Weider[1,11]

[1]Earth and Planets Laboratory, Carnegie Institution of Washington, Washington, DC, USA

[2]School of Earth and Space Exploration, Arizona State University, Tempe, AZ, USA

[3]Geology Department, Department of Physics & Astronomy, Western Washington University, Bellingham, WA, USA

[4]Department of Earth and Planetary Sciences, American Museum of Natural History, New York, NY, USA

[5]Department of Earth and Environmental Sciences, Columbia University, New York, NY, USA

[6]National Museum of Natural History, Smithsonian Institution, Washington, DC, USA,

[7]Astromaterials Research and Exploration Science Division, NASA Johnson Space Center, Houston, TX, USA

[8]Physics Department, The Catholic University of America, Washington, DC, USA

[9]Solar System Exploration Division, NASA Goddard Space Flight Center, Greenbelt, MD, USA,

[10]Physics Institute, University of Bern, Bern, Switzerland

[11]Agile Decision Services, Washington, DC, USA

*Corresponding author, lnittler@asu.edu.

## Abstract

Mercury, the innermost planet, formed under highly reduced conditions, based mainly on surface Fe, S, and Si abundances determined from MESSENGER mission data. The minor element Cr may serve as an independent oxybarometer, but only very limited Cr data have been previously reported for Mercury. We report Cr/Si abundances across Mercury's surface based on MESSENGER X-Ray Spectrometer data throughout the spacecraft's orbital mission. The heterogeneous Cr/Si ratio ranges from 0.0015 in the Caloris Basin to 0.0054 within the high-magnesium region, with an average southern hemisphere value of 0.0008 (corresponding to about 200 ppm Cr). Absolute Cr/Si values have systematic uncertainty of at least 30%, but relative variations are more robust. By combining experimental Cr partitioning data along with planetary differentiation modeling, we find that if Mercury formed with bulk chondritic Cr/Al, Cr must be present in the planet's core and differentiation must have occurred at log $f$O$_2$ in the range of IW-6.5 to IW-2.5 in the absence of sulfides in its interior, and a range of IW-5.5 to IW-2 with an FeS layer at the core-mantle boundary. Models with large fractions of Mg-Ca-rich sulfides in Mercury's interior are more compatible with moderately reducing conditions (IW-5.5 to IW-4) owing to the instability of Mg-Ca-rich sulfides at elevated $f$O$_2$. These results indicate that if Mercury differentiated at a log $f$O$_2$ lower than IW-5.5, the presence of sulfides whether in the form of a FeS layer at the top of the core or Mg-Ca-rich sulfides within the mantle would be unlikely.

## Plain Language Summary

Data returned by NASA's MESSENGER mission, which orbited Mercury from 2011-2015, have shown that the innermost planet formed under highly reducing (relatively low-oxygen) conditions, compared to the other terrestrial planets, but estimates of Mercury's oxidation state are highly uncertain. Chromium, a minor element in planetary materials, can exist in a wide range of oxidation states and its abundance thus can provide information about the chemical conditions under which it was incorporated into rocks. We used data from MESSENGER's X-ray Spectrometer instrument to map the Cr/Si ratio across much of Mercury and found that Cr is heterogeneously distributed. By comparing the average measured Cr abundance to the results of planetary differentiation models (informed by experimental data on how Cr partitions between different phases under different planetary differentiation conditions), we placed new constraints on Mercury's oxidation state and show that further refinement of this quantity could be used to place limits on the presence of sulfides in the planet's deep interior.

## 1. Introduction

Despite the wealth of data returned by the MErcury Surface, Space ENvironment, GEochemistry, and Ranging (MESSENGER) spacecraft during its more than four-year orbital mission, the origin and geological evolution of Mercury remain enigmatic (Solomon et al., 2018). Among MESSENGER's instrument payload suite, the X-Ray Spectrometer (XRS) and Gamma-Ray and Neutron Spectrometer (GRNS) were used to measure and map the surface composition of many geochemically important elements—measurements that reflect both the original starting materials that built Mercury as well as the planet's subsequent geological evolution and impact processes. Data from these instruments revealed that Mercury's crust is enriched in Mg and depleted in Al, Ca, and Fe, relative to other terrestrial planets, and that it is surprisingly rich in volatile elements, including S, Na, K, Cl, and C (Evans et al., 2015; Evans et al., 2012; Nittler et al., 2018; Nittler et al., 2011; Peplowski et al., 2011; Peplowski et al., 2014; Weider et al., 2014). Moreover, maps of elemental abundances and neutron absorption have revealed the presence of several distinct geochemical terranes (Peplowski et al., 2015; Weider et al., 2015; Peplowski & Stockstill-Cahill, 2019), spatially contiguous regions that share a chemical composition distinct from their surroundings. The presence of such terranes most likely reflects crustal formation from partial melting of a chemically heterogeneous mantle (Charlier et al., 2013; McCoy et al., 2018, Namur et al., 2016b).

The MESSENGER XRS detected X-ray fluorescence from the top tens of micrometers of Mercury's surface, induced by incident X-rays emitted from the Sun's corona. The XRS was sensitive to elements with X-ray fluorescent lines in the 1 to 10 keV range, which includes many major and minor rock-forming elements. Global maps constructed from XRS data have been reported for Mg/Si, Al/Si, S/Si, Ca/Si, and Fe/Si ratios (Nittler et al., 2018; Nittler et al., 2020; Weider et al., 2015; Weider et al., 2014). Mg, Al, and Si could be detected under all solar conditions (Weider, et al., 2015), so the Mg/Si and Al/Si maps have complete coverage. In contrast, the heavier elements (e.g., S, Ca, and Fe) could only be detected during solar flares and hence global maps of these elements are incomplete. During the largest flares, it was also possible to detect Ti, Cr, and Mn, although analyses of these elements are more difficult because of their low abundance (<1 wt%), and hence low signal-to-noise ratios. Cartier et al. (2020) recently reported analysis of the full-mission XRS Ti data and argued against the presence of a substantial FeS layer at the base of Mercury's mantle. Our focus here is on Cr, for which only 11 XRS measurements have been reported previously (Weider et al., 2014). As discussed further below, Cr is potentially useful as a probe of redox conditions on Mercury and provides additional information on possible mineral assemblages at the planet's surface.

The high S and low Fe contents observed on Mercury's surface are strong evidence that the planet formed under highly reduced conditions, compared with the other terrestrial planets (McCubbin et al., 2012; Namur et al., 2016a; Nittler et al., 2011; Zolotov et al., 2013). That is, as the availability of O decreases, there is increasing partitioning of S and decreasing partitioning of Fe into silicate melts. Estimates of the oxidation state of Mercury's interior, expressed in terms of the oxygen fugacity, $fO_2$, however, extend over a wide range—from two to seven orders of magnitude below the iron-wüstite (IW) buffer. Additional quantitative constraints on Mercury's oxidation state are therefore greatly needed to better understand the core, mantle, and bulk composition of the planet, its origin, and its geological evolution. For example, recently reported partitioning data (Boujibar et al., 2019) indicate that relating surface K/Th and K/U ratios to bulk volatile abundances (e.g., Peplowski et al., 2011) depends critically on $fO_2$ and on the potential presence of an FeS layer (as suggested by Smith et al., 2012) at the base of Mercury's mantle. In this regard, Cr can potentially be used as an independent oxybarometer as it can occur in a variety of oxidation states and its partitioning behavior depends strongly on valence. For example, during silicate melting $Cr^{2+}$ is more incompatible than $Cr^{3+}$ and hence concentrates in partial melts that may form crustal lava flows (Berry et al., 2006).

We report here a map of Mercury's Cr/Si ratio, with partial coverage across the globe, based on XRS spectra acquired during large solar flares throughout MESSENGER's orbital mission. We find that Cr is heterogeneously distributed and correlates with other geochemical parameters. We further use the measured surface Cr abundance along with a large body of experimental data on Cr partitioning between silicates, sulfides, and metal together with planetary differentiation modeling to investigate the conditions under which Mercury differentiated and assess the redox conditions under which the innermost planet differentiated. In addition, we tested scenarios where sulfides are involved in Mercury's differentiation to further address the distribution of Cr in its interior.

## 2. MESSENGER XRS Data Processing

We derived elemental abundances from MESSENGER XRS spectra through an iterative forward modeling/non-linear curve-fitting procedure in which the abundances themselves are fit parameters (Nittler et al., 2011). An example spectrum acquired during a very large X-class solar flare on 23 October 2012 is shown in Figure 1. This spectrum is the sum of spectra from the three individual gas proportional counters that made up the planet-facing portion of the XRS, summed over 11 individual 20-s XRS integrations. Fitting of the solar spectrum (not shown) simultaneously acquired by the Sun-pointing solar monitor indicated a very high solar coronal temperature of >30 MK. Unlike Mg, Al, Si, Ca, S, and Fe, all of which show peaks in the spectrum (although unresolved from each other in many cases), the low abundance of Cr means that its fluorescent photons do not form a distinct peak, but rather contribute to the continuum between the Ca and Fe $K_\alpha$ lines. Nevertheless, during large flares like that shown in Fig. 1, the signal-to-noise ratio in the XRS spectra is sufficient that the Cr abundance can be constrained well by the fitting procedure. In Fig. 1, the fitting procedure is indicated by the grey curves, which compare the best-fit Cr abundance with spectra corresponding to higher and lower abundances. We note that although the summed detector spectra are shown in Fig. 1, the actual fitting algorithm fits the three individual detector spectra individually.

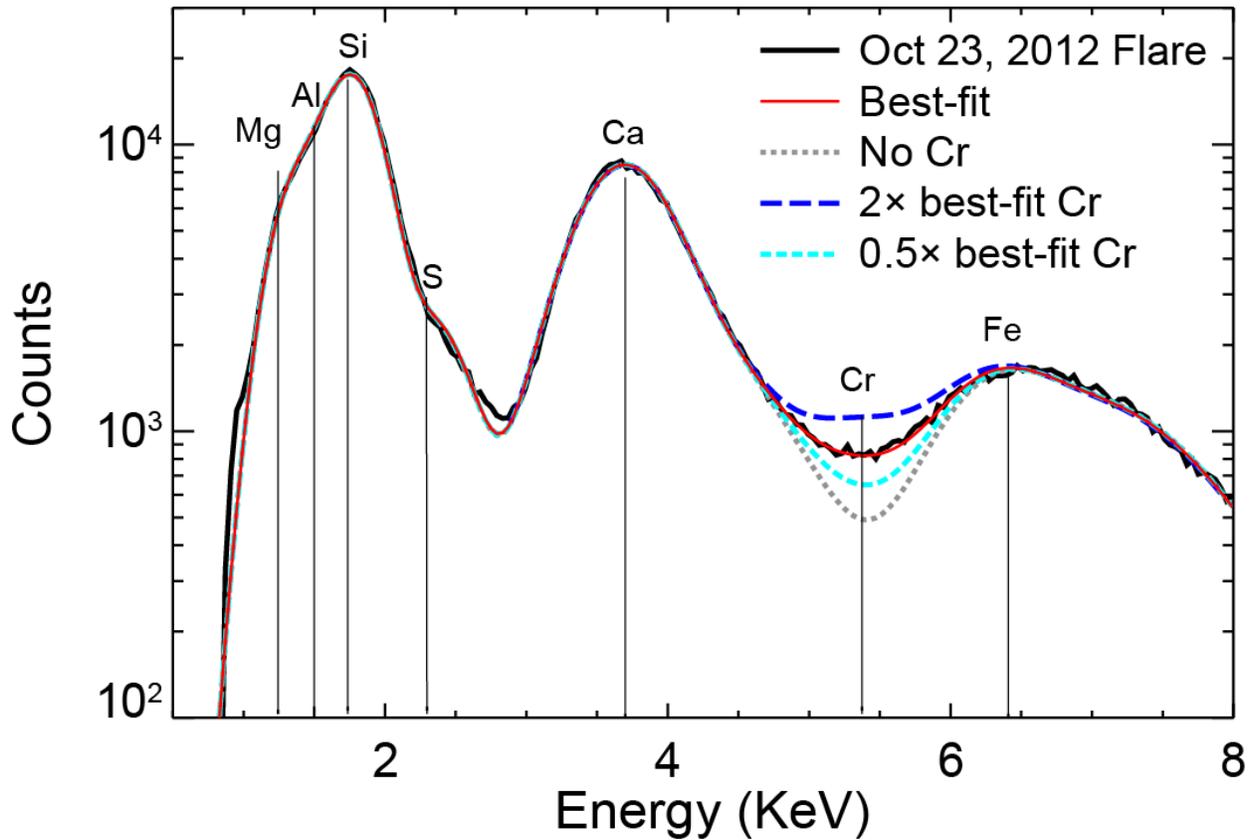

**Figure 1.** Example MESSENGER XRS data. Summed XRS spectra (sum of three detectors) acquired during a large solar flare on October 23, 2012. Included spectra span Mission Elapsed Time (MET) 259449322 to 259449622, corresponding to 3:10:56 – 3:14:16 UTC.

A total of 2300 spectral fits to XRS data from MESSENGER's full orbital mission, acquired during 291 distinct solar flares, were performed—including fits to individual integrations during flares and to spectra summed over entire flare periods (Nittler et al., 2020). Cr abundance was included as a fit parameter for all the data for which Fe fluorescence was clearly observed, although in many cases the signal was too low to detect Cr and the fitting procedure returned a best-fit abundance of zero. We examined the full dataset and selected results from 133 flare spectra from which to construct our Cr/Si map. Each of the spectra we selected had a derived Cr abundance >0 and did not exhibit anomalously high detector backgrounds at high energy. It has previously been shown that such high detector backgrounds arose during some flares because of interactions between solar charged particles and the XRS detectors (Weider et al., 2014) and that they contaminate XRS measurements in the Cr fluorescence region.

In Weider et al. (2014), the measured Cr/Si and Fe/Si ratios were empirically corrected for an observed dependence on phase (Sun-planet-instrument) angle $\phi$; this dependence is thought to be caused by shadowing effects on the planet's non-flat surface. Our larger data set confirms this effect for Cr/Si, as illustrated in Figure 2. Filled circles in Fig. 2 indicate the measured Cr/Si ratios as a function of phase angle for 24 flare fits with high statistical significance and large footprints (projections of the XRS instrument's field of view onto the planet's surface) in the southern hemisphere. Because of the very poor spatial resolution of the XRS over the southern hemisphere, these footprints are assumed to all have the same true Cr/Si ratio. We therefore removed the phase-angle dependence from all the flare measurements by dividing the measured Cr/Si ratio by the ratio predicted for its phase angle from the best-fit line to the southern hemisphere data (solid line in Fig. 2). This procedure, however, introduces an overall ambiguity in the overall normalization.

Several theoretical and experimental studies addressing particle and shadowing effects on remote XRF measurements of planetary surfaces have been published (e.g., Maruyama et al., 2008; Weider et al., 2011; Parviainen et al., 2011). Although none of these considered phase angles higher than 80°, i.e., as seen for much of the MESSENGER XRS data set, several did consider angles in the 70-80° range and can thus be compared with the low-angle end of the trend in Fig. 2. Based on laboratory experiments, Maruyama et al. (2008) presented numerical estimates of the phase-angle effect on XRF line intensities, with an assumed solar flare incident spectrum, lunar soil composition, and 75-$\mu$m grain size. At the maximum phase-angle they considered, 75°, they found that the measured Ti/Si and Fe/Si ratios are higher than those predicted for a flat surface by factors of about 1.4 and 1.6, respectively (values were estimated from data plotted in their Figure 7). Weider et al. (2011) reported an effect of very similar magnitude for Fe/Si ratios that were measured by irradiating simple oxide mixtures with an X-ray beam generated from a Cu anode and phase angles of 70-80° (see their Fig. 12). Parviainen et al. (2011) used Monte Carlo ray-tracing calculations to investigate geometric effects on XRF from a basalt composition as a function of a variety of particle sizes, porosities, and incident spectra. For a polychromatic incident X-ray source, they found enhanced Ti/Si fluorescence line ratios ranging from ~1.1 to 1.5 at a phase angle of 75° (see their Fig. 6). The $K_\alpha$ XRF line of Cr lies between those of Ti and Fe. We thus adopt an enhancement factor for Cr/Si at $\phi$ =75° of 1.5, in between the Maruyama et al. (2008) values for Ti/Si and Fe/Si since this study is the most relevant physical analog to the MESSENGER data (in terms of assumed incident spectrum and composition).

The fit line to the data on Fig. 2 provides a measured Cr/Si ratio across the southern hemisphere of Mercury of 0.00112 at $\phi$ =75°. The discussion above indicates that this must be divided by a factor of 1.5 to determine the intrinsic average Cr/Si ratio of $8 \times 10^{-4}$. We thus renormalized the data set so that the average Cr/Si value in the southern hemisphere was equal to this value; the corrected data for the 24 southern hemisphere flares are shown as open triangles in Fig. 2. This average value is ~9 times lower than the value of 0.007 adopted by Weider et al. (2014). Based on the scatter in experimental/theoretical works discussed above, we estimate that the overall relative uncertainty in this average value is about 30%, or $2.4 \times 10^{-4}$. Systematic uncertainties may well be larger but are difficult to assess. Even if this is the case, relative differences between our mapped Cr/Si values, however, are much more certain than the absolute normalization. The relative scatter in the corrected Cr/Si values for the southern hemisphere measurements, where the large XRS footprints overlap a great deal, is ~30% (one standard deviation), indicating that the flare-to-flare reproducibility of the Cr/Si measurements is no worse than this (e.g., some of this variability may reflect real large-scale heterogeneity across the southern hemisphere).

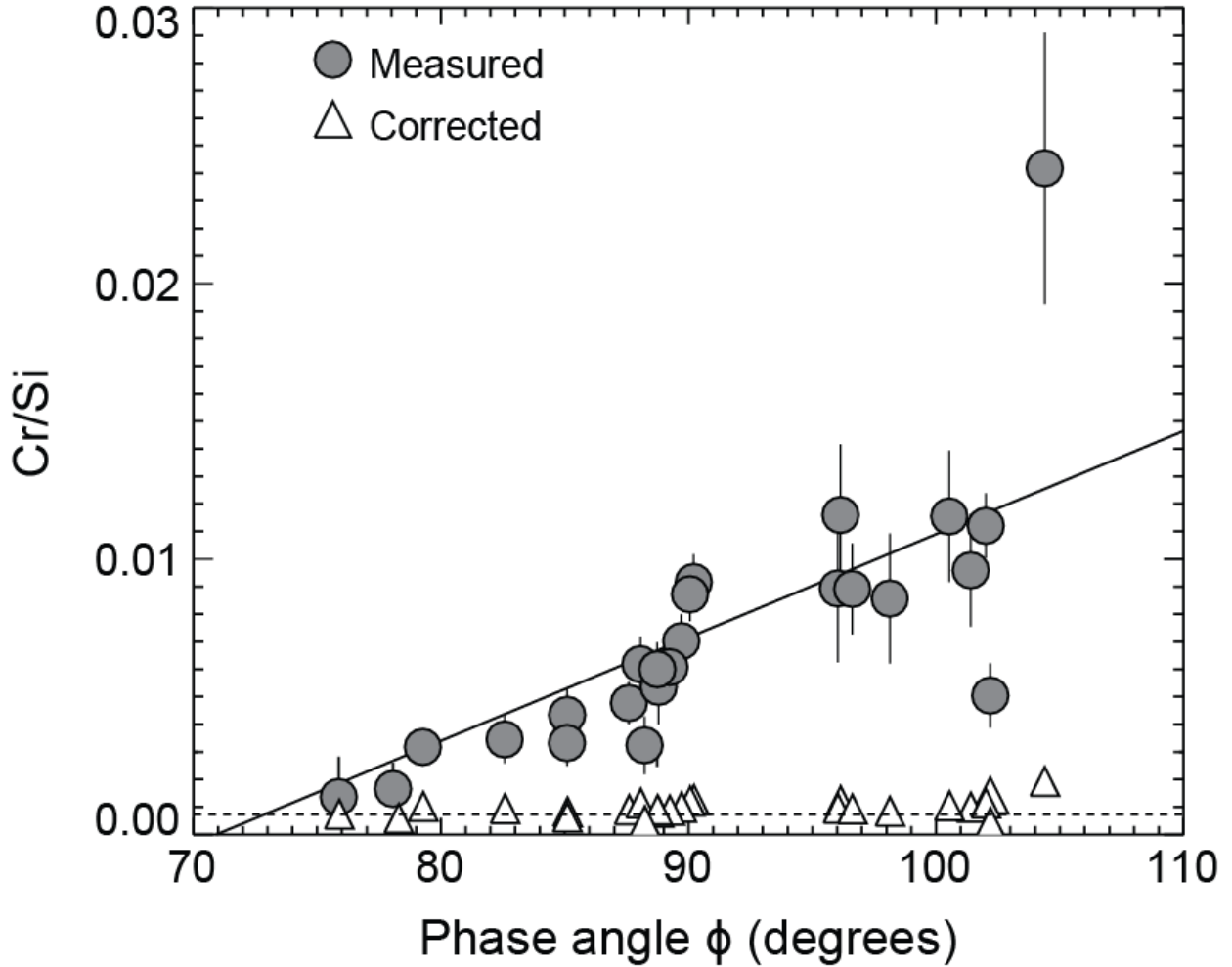

**Figure 2.** Phase-angle effect on Cr/Si ratios. Measured Cr/Si ratios (filled circles) are plotted as a function of phase angle for 24 XRS measurements that have footprints in Mercury's southern hemisphere and relatively small error bars. A weighted linear fit to the data (solid line) gives Cr/Si = 0.000336 $\phi$ – 0.024. Removing this trend and renormalizing the data to an average value of 0.0008 (dashed line) yields the corrected Cr/Si values (open triangles).

We used the same procedure we have used in previous work (Nittler et al., 2020; Weider et al., 2015; Weider et al., 2014) to generate a Cr/Si map from the 133 flare measurements and corresponding footprints. Briefly, the surface was divided into 0.25°×0.25° pixels in cylindrical projection. The Cr/Si value of a given pixel in the map is based on a weighted average of all measurements that have footprints overlapping that pixel, with a weighting factor favoring measurements with smaller footprints and statistical uncertainty. The map was further smoothed following the procedure described by Weider et al. (2015) and Nittler et al. (2020).

## 3. The Surface Abundance of Cr on Mercury

Our map of Cr/Si across Mercury's surface is shown in Figure 3a. The sparsity of data for the northern hemisphere (where XRS spatial resolution is best) is the result of two factors. First, MESSENGER's highly eccentric polar orbit meant that far more time was spent viewing the southern hemisphere than the northern hemisphere of the planet. Second, as described above, large flares were required for Cr detection, but such flares occur infrequently. Nonetheless, there is clear evidence for heterogeneity in Cr/Si across Mercury's surface. The XRS maps of Mg/Si and Al/Si (Nittler et al., 2018) are also shown, for comparison, in Figs. 3b and c. The white lines indicate the locations of the high-magnesium region (HMR) and Caloris basin (CB) geochemical terranes, and the northern smooth volcanic plains (NSP). There is too little coverage for Cr/Si in the NSP to make meaningful comparisons with other regions, but the map includes multiple flare measurements across the HMR and CB. These two terranes represent compositional endmembers on Mercury (Nittler et al., 2018), i.e., the HMR has the highest Mg/Si, S/Si, Ca/Si, and Fe/Si and lowest Al/Si ratios on Mercury, whereas the opposite trends are true of the CB. Fig. 3a suggests that, similarly, the HMR and CB have higher and lower average Cr/Si ratios, respectively, than the average planetary value.

To investigate the Cr/Si heterogeneities further, we generated histograms (Fig. 4a) of Cr/Si map pixel values within the HMR (only considering pixels north of 15°N latitude since the spatial resolution rapidly degrades southward of this), CB, and the average Mercury composition outside these terranes (intermediate terrane, IT). The histograms are weighted to favor pixels with higher spatial resolution (Nittler et al., 2020). Although we find that there is overlap between the IT and HMR histograms, largely due to large errors on individual measurements, there are significant differences in the average composition of each terrane. That is, the HMR has an average Cr/Si ratio that is 1.5±0.5 times the IT average, and the CB average is 0.45±0.02 times that of the IT (errors are one standard deviation). Thus, like other major and minor elements, the Cr abundance varies across Mercury's surface and appears to correlate with the Mg, S, Ca, and Fe abundances, and anti-correlate with that of Al, at least in large geochemical terranes (Fig. 4b). Unfortunately, there are almost no Cr measurements on the northern smooth plains, which have previously been shown to have a range of chemical compositions (Lawrence et al., 2017; Weider et al., 2015).

The major-element heterogeneity on Mercury's surface is generally considered to reflect partial melting of a heterogeneous mantle. As discussed in Section 4.2, Cr is incompatible in pyroxene at the reducing conditions (log $f$O$_2$ <IW) inferred for Mercury, but experimental data for olivine-melt partitioning of Cr under such conditions have not been reported. The correlation of Cr with compatible elements Mg and Ca and anti-correlation with incompatible Al across Mercury's surface suggests that either Cr remains compatible with olivine under low $f$O$_2$ conditions or that sulfides fractionate Cr in Mercury's mantle. One explanation of this trend could be the presence of more sulfides closer to the surface (e.g., the Caloris basin magma's source region), since sulfide solubility in silicate melt decreases when decreasing temperature (e.g. Namur et al. 2016a). Closer to the surface, Ca-Mg-Fe-rich sulfides could deplete the silicates in S, Cr, Ca, Fe and Mg. Additional experimental data are needed to further investigate the origin of the observed elemental trends.

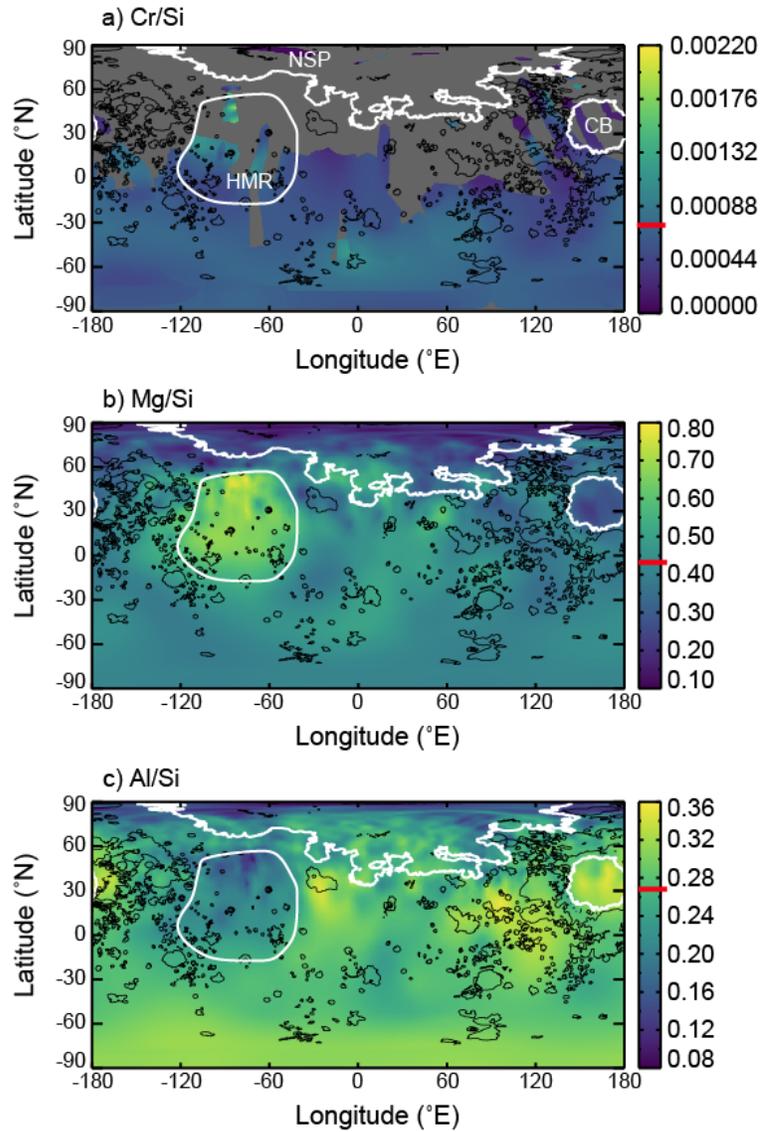

**Figure 3.** Maps of a) Cr/Si, b) Mg/Si, and c) Al/Si across Mercury's surface (shown in cylindrical projection). The white outlines indicate previously identified features: CB=Caloris Basin, NSP=Northern smooth plains (Head et al., 2011), HMR=high Mg region (Weider et al., 2015). Smooth plains deposits (Denevi et al., 2013) are outlined in black. Global average values are indicated by red lines on color bars.

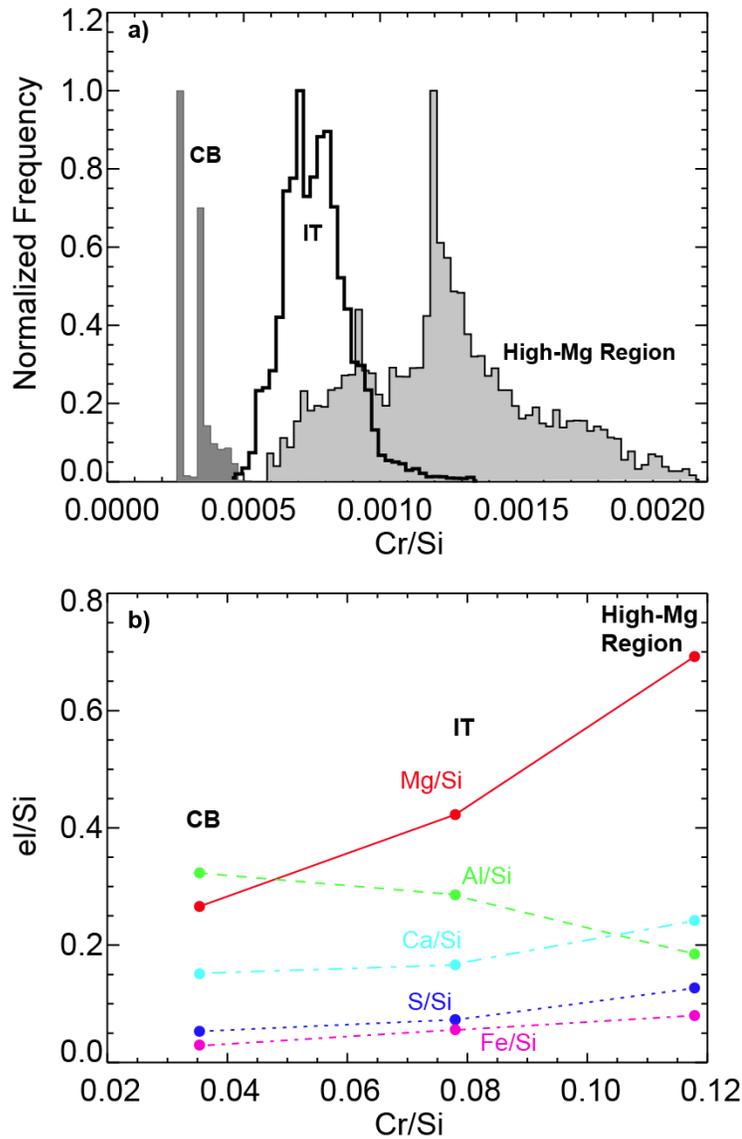

**Figure 4.** a) Histograms of Cr/Si ratios within CB, HMR, and intermediate terrane (IT), the latter defined as the composition of the southern hemisphere. The histograms are individually scaled since they contain vastly different numbers of pixels. b) The average element/silicon ratios plotted versus Cr/Si ratios for Mg, Al, S, Ca, and Fe for the CB, HMR, and IT terranes.

The average Cr/Si value of the map is forced by our assumed normalization scheme to be close to 8×10⁻⁴. The MESSENGER XRS element-to-silicon ratios can be converted into absolute elemental composition by assuming a valence state for the major cations and adding the appropriate amount of oxygen (Lawrence et al., 2013; Stockstill-Cahill et al., 2012; Vander Kaaden et al., 2017). Such calculations have resulted in estimated absolute Si abundances of ~24–27 wt%. If we assume a typical Si abundance of 25 wt%, we estimate the average surface Cr abundance to be 200 ppm. In turn, this implies average Cr abundances of $300\pm100$ ppm and $90\pm4$ ppm in the HMR and Caloris basin, respectively. These uncertainties are based on the standard deviations of the image pixels in each region. As discussed above, the phase-angle correction introduces an additional relative systematic uncertainty of at least 30% (e.g., 60 ppm for IT).

## 4. Cr partitioning and planetary differentiation modeling

### 4.1. Cr behavior in planetary materials

The geochemical behavior of multivalence elements is dependent on oxygen fugacity and the oxidation state of Cr has long been recognized as a potentially useful oxybarometer for planetary basalts (e. g.,Irvine, 1975; Papike, 2005). Of the three most common valence states for Cr ($Cr^0$, $Cr^{2+}$, and $Cr^{3+}$), $Cr^{2+}$ is the most geochemically incompatible, i.e., it is the most easily liberated by minerals into melt during partial melting. Experiments on terrestrial (Berry et al., 2006, Righter et al. 2016) and martian (Bell et al., 2014) basaltic compositions have shown that as $fO_2$ decreases, the ratio of divalent Cr to total Cr ($Cr^{2+}/\sum Cr$) increases in basaltic liquids from ~0.3–0.5 at oxygen fugacity similar to mid-ocean ridge basalts (MORB; $\log fO_2$ ~IW+3.5) to ~0.8–0.9 at lunar-like oxygen fugacity ($\log fO_2$ near IW–1). In addition, Cr content of basalts range from typical MORB values of $250 \pm 165$ ppm (Lehnert et al., 2000) to $5200 \pm 700$ ppm for lunar basalts (Delano, 1986). Although Cr incompatibility reaches a theoretical maximum at $Cr^{2+}/\sum Cr = 1$, with further drop in $fO_2$, Cr becomes chalcophilic and is more likely to occur in sulfides than silicates (Vander Kaaden & McCubbin, 2016). Moreover, the stability of $Cr^0$ increases as the $fO_2$ approaches that of the Cr-$Cr_2O_3$ buffer. As a result, although the mantle-melt partition coefficient for Cr continues to decrease with decreasing $fO_2$, much of the Cr may go into a metallic phase that could segregate to the core and thus reduce the bulk Cr content of surface lavas. The aubrites—highly reduced achondrites, thought to have equilibrated at approximately IW-5 (McCoy & Bullock, 2017) — have average Cr abundances of only 200 ppm (Keil, 2010), consistent with loss of $Cr^0$ to metallic melts. These changes in Cr behavior with oxygen fugacity are illustrated in Figure 5. The measured Mercury surface Cr abundance reported here, 200($\pm$60) ppm is very similar to aubrites, suggesting that Mercury has a similar $\log fO_2$ close to IW-5 (Fig. 5). However, these estimates do not take into account the possible presence of an FeS sulfide layer at the core-mantle boundary of Mercury (Smith et al., 2012) or the possible presence of Mg-Ca-rich sulfides in the mantle of Mercury (Namur et al., 2016a), both of which affect the overall bulk distribution coefficient for Cr (see below sections 4.2-4.3) (Steenstra et al., 2020; Vander Kaaden & McCubbin, 2016).

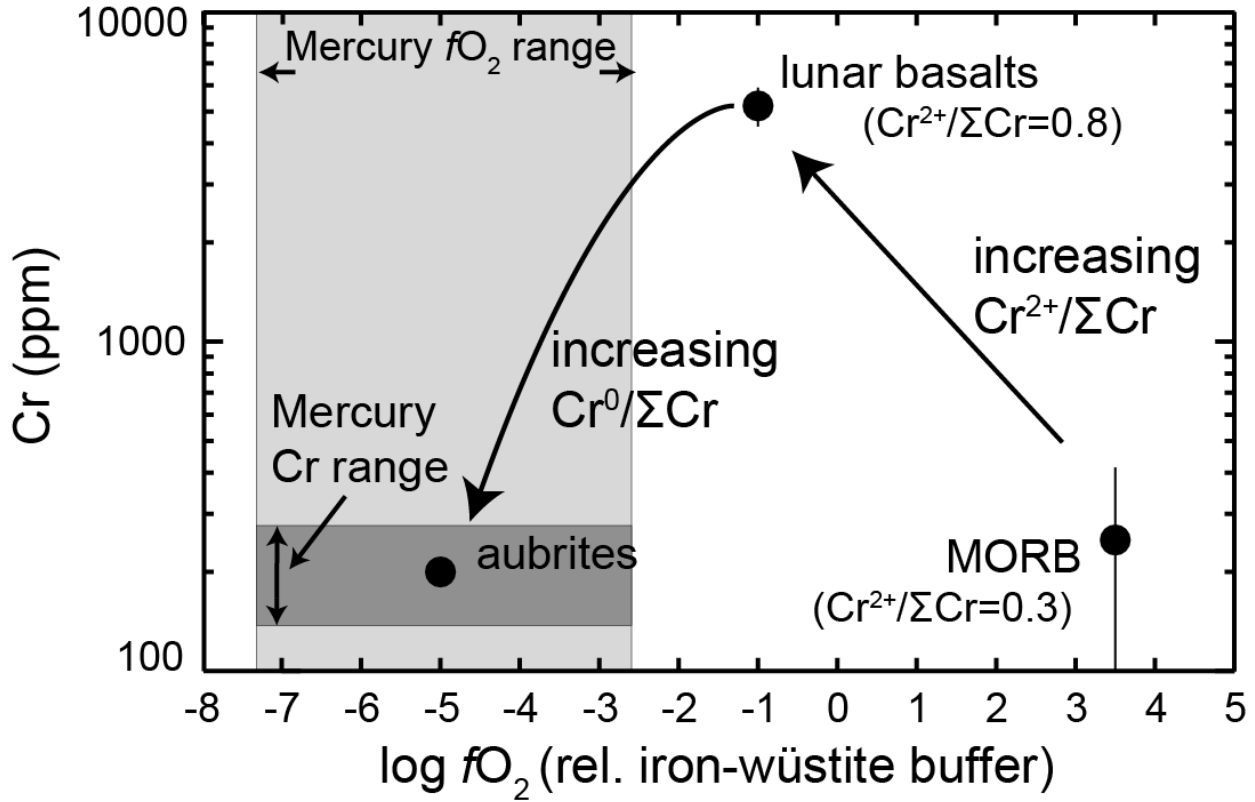

**Figure 5.** Effect of oxygen fugacity on Cr abundance in planetary materials. Filled circles indicate Cr abundances and estimated oxygen fugacities for mid-ocean ridge basalts (MORB), lunar basalts, and aubrite meteorites (see text for references). Also indicated are estimated ratios of divalent Cr to total Cr for MORB and lunar samples (Berry et al. 2006; Bell et al. 2014) as well as the range of estimated oxygen fugacity for Mercury (see text) and estimated range of Cr/Si across Mercury's surface from this work.

Mercury's overall bulk composition is known to be strongly non-chondritic, with a substantially higher abundance of iron indicated by its unusually large core (Nittler et al., 2018). Moreover, depending on the amount of Si in the core, Mercury's bulk Si may also be enriched relative to chondritic. However, other major elements like Mg, Al, and Ca do appear to be in chondritic proportions in Mercury (Nittler et al., 2018). To constrain the planet's oxidation state during its differentiation more quantitatively, we thus assume Mercury's bulk Cr abundance is chondritic and use experimental partitioning data and differentiation modeling to investigate conditions under which this assumption is valid. We focus on the Cr/Al ratio, since Al is a refractory lithophile element and therefore is only weakly fractionated in chondrites. Based on an average Al/Si ratio of 0.27 (Nittler et al., 2018), the average surface Cr/Al ratio for Mercury is $0.003 \pm 0.001$, which is considerably lower than that of chondritic meteorites (ranging from 0.05 to 0.5; Nittler et al., 2004). We used surface abundances to estimate bulk silicate Mercury (BSM) Cr content, which is assumed to represent the magma ocean composition during core–mantle differentiation. We then used the partitioning of Cr between metal and silicate to model core composition and hence infer bulk Mercury composition. We further assessed differentiation models by considering the presence of an additional sulfide phase formed during core formation or magma ocean crystallization, i.e., because of the immiscibility of Fe-rich metal and Fe-Mg-Ca-rich sulfides, respectively.

## 4.2. Experimental constraints on Cr partitioning between metal, sulfide, silicate melt, and minerals

Our planetary differentiation models use experimental data on elemental partitioning. The partitioning of Cr between minerals and silicate melt can be described by the partition coefficient: $D_{Cr}^{mnl/melt} = X_{Cr}^{mnl}/X_{Cr}^{melt}$, where $X_{Cr}^{mnl}$ and $X_{Cr}^{melt}$ are wt% concentrations of Cr in minerals and silicate melt respectively. Experimental studies have shown that under moderately oxidized conditions (IW $< \log fO_2 <$ IW+9) where $Cr^{6+}$ is absent, $D_{Cr}^{ol/melt}$ (ol: olivine) has a relatively constant value of $0.9 \pm 0.3$ (Hanson & Jones, 1998; Mallmann & O'Neill, 2009). In contrast, $D_{Cr}^{opx/melt}$ and $D_{Cr}^{cpx/melt}$ (opx: orthopyroxene, cpx: clinopyroxene) increase by an order of magnitude when oxygen fugacity increases from IW to IW+9, suggesting a preferential incorporation of $Cr^{3+}$ over $Cr^{2+}$ in pyroxenes (Mallmann & O'Neill, 2009). As discussed above in Section 1, Mercury differentiated at very low oxygen fugacity with estimates covering a wide range from IW–7 to IW–2.6 (McCubbin et al. 2012, McCubbin et al. 2017, Namur et al. 2016a, Zolotov et al. 2013). Previous experimental work showed that in these reduced conditions, $D_{Cr}^{opx/melt}$ ranges from 0.1 to 0.6 (Cartier et al. 2014). The partitioning of Cr between olivine/clinopyroxene and liquid silicate in these specific conditions, however, is unknown and should be investigated in future studies. Here, we considered a distribution of Cr between mantle and crust of 0.35 ± 0.25, covering values determined experimentally for orthopyroxene/melt in Mercury conditions (Cartier et al. 2014).

Next, we modeled the distribution of Cr between Mercury's core and BSM based on experimental data for Cr partition coefficient between metal and silicate $D_{Cr}^{met/sil} = X_{Cr}^{met}/X_{Cr}^{sil}$, where $X_{Cr}^{met}$ and $X_{Cr}^{sil}$ are concentrations in wt% of Cr in the metal and silicate liquids, respectively. $D_{Cr}^{met/sil}$ can be described by the redox reaction:

$$CrO_{n/2}{}^{sil} = Cr^{met} + \frac{n}{4}O_2 \qquad (1)$$

This reaction implies that the change of Cr partitioning with oxygen fugacity depends on $n$, the valence state of Cr in silicates. The equilibrium constant of this reaction can be related to its free energy $\Delta G°$:

$$-\Delta G°/RT = \frac{n}{4}\ln fO_2 + \ln a_{Cr^{met}} - \ln a_{CrO_{n/2}{}^{sil}} \qquad (2)$$

Where $a_{Cr^{met}}$ and $a_{CrO_{n/2}{}^{sil}}$ are the activities of Cr in the metal and $CrO_{n/2}$ in the silicate, respectively. Using a common formalism for the dependencies of activity coefficients with chemical compositions (e.g., Boujibar et al., 2019), we derived the following expression to calculate $D_{Cr}^{met/sil}$ as a function of pressure P, temperature T, the logarithm of oxygen fugacity relative to the iron-wüstite buffer (ΔIW) and the chemical composition of the silicate and metal phases:

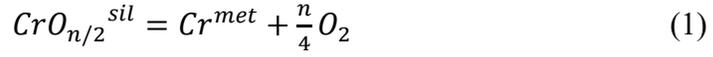

$$\log D_{Cr}^{met/sil} = \log\frac{X_{Cr}^{met}}{X_{Cr}^{sil}} = a + \frac{b}{T} + \frac{c*P}{T} + d*\Delta\text{IW} + e*\frac{T_0\log(1-X_{Si}^{met})}{T} + f*\frac{T_0\log(1-X_S^{met})}{T} + g*$$

$$\frac{T_0\log(1-X_C^{met})}{T} + h*\frac{T_0\log(1-X_O^{met})}{T} + i*nbo/t \qquad (3)$$

$X_M^{met}$ are mass fractions of light elements M (Si, S, O, and C) in the metal alloy, $T_0$ is a reference temperature (1873 K), and nbo/t is the ratio of non-bridging O atoms to tetrahedrally coordinated cations, which carries effects of the chemical composition of the silicate melt. Experiments using a graphite capsule are known to yield contamination of the metal and sulfide by carbon. Therefore, if these experiments were not measured for C concentration in the metal and sulfide phases, we computed C abundance by subtracting the sum of all measured elemental concentrations from 100% (see supplementary material.) We calculated the logarithm of the oxygen fugacity relative to IW buffer from the following equation: $\Delta IW = 2 * \log(\gamma_{FeO}^{sil} * X_{FeO}^{sil}) - \log(X_{Fe}^{met})$ where $X_{FeO}^{sil}$ and $X_{Fe}^{met}$ are the molar fractions of FeO in the silicate and Fe in the metal, respectively. $\gamma_{FeO}^{sil}$ is the activity coefficient of FeO in the silicate and is considered equal to 1.7 following previous estimates (O'Neill & Eggins, 2002).

A substantial amount of experimental data exist for the partitioning of Cr between metal and silicate; here we used 520 experimental data from 43 peer-reviewed publications whose references are given in the supplementary material. These data on $D_{Cr}^{met/sil}$ at varying experimental conditions were used to derive constants a to i in Eq. (3) using a linear regression (see results in Table 1). Equation (2) informs that at constant pressure and temperature, $\ln(a_{Cr^{met}}/a_{CrO_{n/2}^{sil}})$ should be proportional to $\ln fO_2$ with the activities (*a*) being the products of the activity coefficients and the molar mass fractions. The effects of the activity coefficients of Cr in the silicate and the metal are included in our thermodynamic model (Eq. 3) in the nbo/t term and the $\log(1 - X_M^{met})$ terms, respectively. For the sake of simplicity in our models of Mercury's differentiation, we used Nernst partition coefficients $D_{Cr}^{met/sil}$, which are calculated using elemental concentrations in wt%. Since $D_{Cr}^{met/sil}$ and $X_{Cr}^{met}/X_{Cr_2O_3}^{sil}$ are proportional (see Supplementary Fig. S1), the use of a logarithm expression allows the effect of molar to weight fraction conversion to be included in the constant a of equation (3). Hence, $D_{Cr}^{met/sil}$ is expected to be correlated with $\log fO_2$ with a slope equivalent to n/4, with n being the valence of Cr in the silicate melt. Here, we observe a negative correlation with $\log fO_2$, with a slope of -0.52 ± 0.01, suggesting Cr predominantly has a valence state of $2^+$ in the silicate melt at these conditions (Fig. 6b). The pressure term is found to be insignificant (p-value higher than 10%). In addition, $D_{Cr}^{met/sil}$ becomes more siderophile with increasing temperature (Fig. 6b), as previously reported (e.g., Fischer et al., 2015; Righter et al., 2020) and with increasing abundances of Si, S, C and O in the metal. The nbo/t ratio has a negative effect on Cr partitioning between metal and silicate (Table 1).

**Table 1.** Fitted parameters for linear regressions predicting the partition coefficient of Cr between metal and silicate (a to i) (Eq. 3), and between sulfide and silicate (a' to i') (Eq.5).

| Metal-silicate | a [intercept] | b [1/T] | d [$\Delta$IW] | e [log(1-$X_{Si}$)/T] | f [log(1-$X_S$)/T] | g [log(1-$X_C$)/T] | h [log(1-$X_O$)/T] | i [nob/t] |
|---|---|---|---|---|---|---|---|---|
| Coef | -0.15 | -2870 | -0.52 | -1.61 | -4.96 | -12.9 | -34.8 | -0.041 |
| $\sigma$ | 0.16 | 290 | 0.01 | 0.36 | 1.24 | 1.4 | 6.58 | 0.017 |
| P-value | 0.4 | <2e-16 | <2e-16 | 1E-5 | 8E-5 | <2e-16 | 2E-7 | 2E-2 |
| N | 520 | RMSE | 0.33 | $R^2$ | 0.84 | F | 386 | |
| P-value | <2e-16 | | | | | | | |

| Sulfide-silicate | a' [intercept] | b' [1/T] | d' [$X_{FeO}$] | e' [log(1-$X_S$)/T] | f' [log(1-$X_C$)/T] | g' [log(1-$X_O$)/T] | h' [log(1-$X_{Mg}$)/T] | i' [nob/t] |
|---|---|---|---|---|---|---|---|---|
| Coef | 3.9 | -8370 | -0.89 | -11.9 | | -27.9 | 11.6 | -0.37 |
| $\sigma$ | 0.46 | 950 | 0.05 | 1.6 | | 6.1 | 1.6 | 0.06 |
| P-value | 2e-15 | <2e-16 | <2e-16 | 5E-12 | | 9e-6 | 2E-12 | 8E-9 |
| N | 253 | RMSE | 0.46 | $R^2$ | 0.76 | F | 127 | |
| P-value | <2e-16 | | | | | | | |

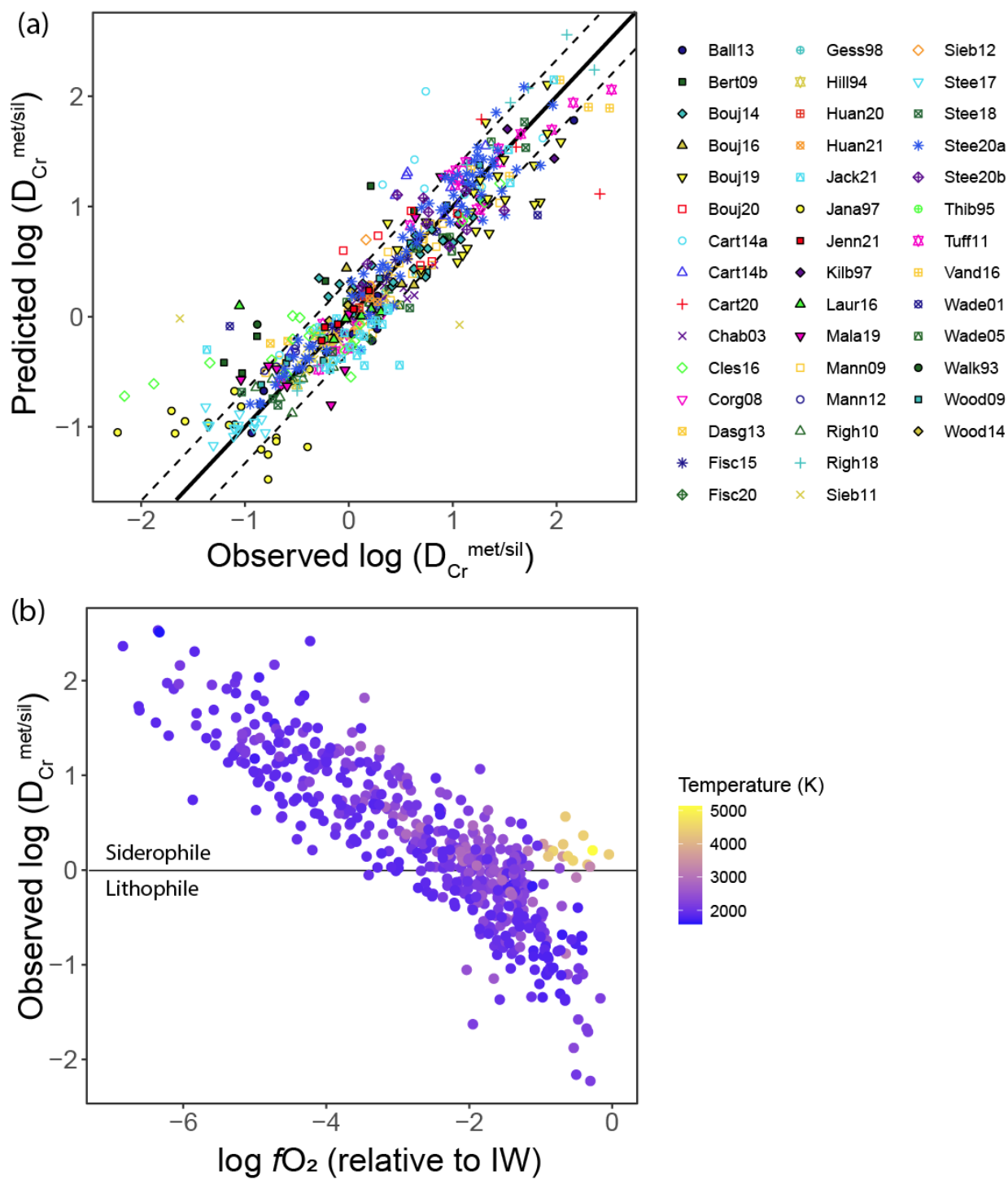

**Figure 6.** (a) Experimental data on the partition coefficients of Cr between metal and silicate compared with predictions from our thermodynamic model (Eq. 3). The experimental data are derived from the literature, and the complete list of references is given in the supplementary

material. The solid and dashed lines represent the 1:1 correspondence and deviation, respectively, where the deviation is based on the RMSE (=0.33) from our regression. (b) Relationship between $D_{Cr}^{met/sil}$ and (i) the oxygen fugacity and (ii) temperature (shown with symbol color) based on the same experimental data presented in (a). The same data for panel b, but with literature references indicated, is provided in Supplementary Fig. 2a.

In addition, we explored the possibility of the existence of an immiscible sulfide formed during core-mantle differentiation by using the equilibrium:

$$CrO_{n/2}{}^{sil} + \frac{n}{2}FeS = CrS_{n/2}{}^{sulf} + \frac{n}{2}FeO \qquad (4)$$

Similarly to reaction (1), we related the equilibrium constant of reaction (4) with its free energy to construct a thermodynamic model that predicts the partition coefficient of Cr between sulfide and silicate $D_{Cr}^{sulf/sil} = X_{Cr}^{sulf}/X_{Cr}^{sil}$, where $X_{Cr}^{sulf}$ is the concentration (wt%) of Cr in the sulfide in (Boujibar et al. 2019):

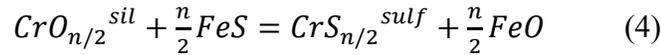

$$\log D_{Cr}^{sulf/sil} = \log \frac{X_{Cr}^{sulf}}{X_{Cr}^{sil}} = a' + \frac{b'}{T} + \frac{c'*P}{T} + d'*X_{FeO} + e'*\frac{T_0 \log\left(1-X_S^{sulf}\right)}{T} + f'*\frac{T_0 \log\left(1-X_C^{sulf}\right)}{T} + g'*\frac{T_0 \log\left(1-X_O^{sulf}\right)}{T} + h'*\frac{T_0 \log\left(1-X_{Mg}^{sulf}\right)}{T} + i'*nbo/t$$

$$(5)$$

$X_S^{sulf}, X_C^{sulf}, X_O^{sulf}$ and $X_{Mg}^{sulf}$ are the concentrations of S, C, O and Mg in the sulfide, respectively. Equation 5 was fit to a total of 253 experimental data from the literature (see complete list of reference in the supplementary material), in which sulfides have varying compositions (from Fe-rich to Ca-Mg-rich compositions). We did not find any significant effects from pressure, or from the C or Ca contents of the sulfide. In contrast, we found that temperature and the abundances of S and O in the sulfide have a positive effect on $D_{Cr}^{sulf/sil}$ and that FeO, nbo/t, and Mg content in the sulfide attenuates the chalcophilic character of Cr (Fig. 7b & Table 1).

(a)

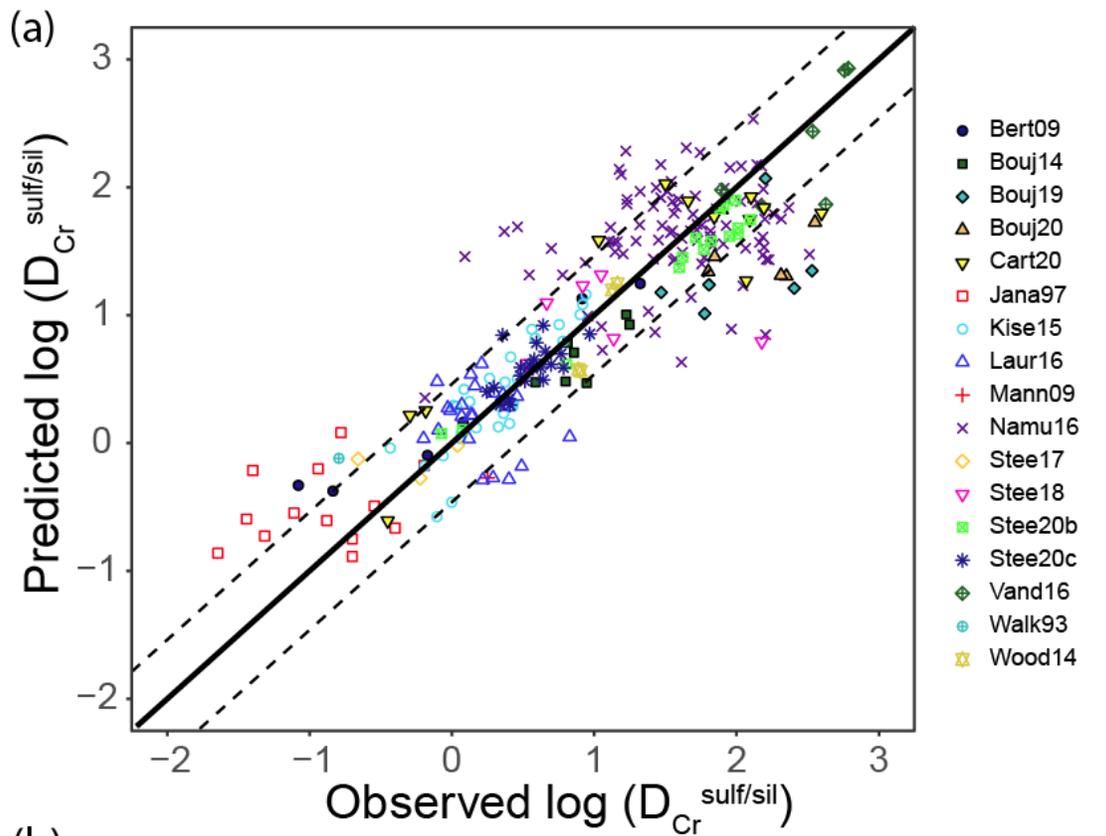

(b)

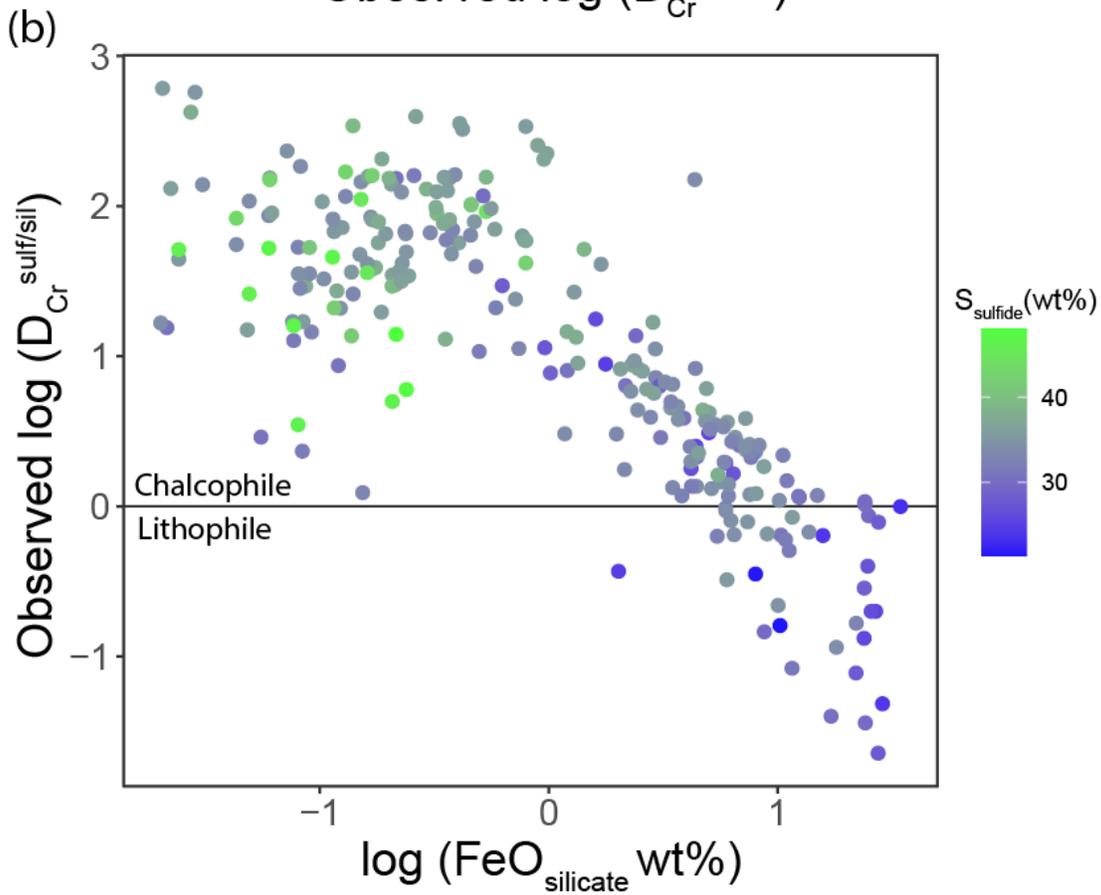

**Figure 7.** (a) Comparison between observed experimental data and predicted values for $D_{Cr}^{sulf/sil}$, i.e., results of our thermodynamic model (Eq. 5). The solid and dashed lines represent the 1:1 correspondence and deviation, respectively, based on the RMSE of our regression (RMSE = 0.46). (b) Relationship between $D_{Cr}^{sulf/sil}$ and FeO concentration in the silicate and S contents of the sulfide phase (indicated by plot symbol color). The same data for panel b, but with literature references indicated, is provided in Supplementary Fig. 2b.

### 4.3. Modeling Mercury's differentiation:

We used the average XRS-derived Cr concentration of Mercury's surface (Section 3), along with chondritic Cr abundances normalized to Al (Cr/Al ranges from 0.0472 to 0.509 for ordinary, enstatite, carbonaceous and R chondrites; Nittler et al. 2004) and the thermodynamic model results for Cr partitioning between metal/sulfide and silicate (Section 4.2), to estimate the Cr concentration of Mercury's interior and to infer the oxygen fugacity at which Mercury differentiated. First, we assumed that the average crustal Cr concentration is close to the average surface concentration, and calculated a bulk silicate Mercury (BSM) composition using the literature data on Cr partitioning between silicate melt and major minerals (mnl being olivine, orthopyroxene, or clinopyroxene): $D_{Cr}^{mnl/melt} = X_{Cr}^{mnl}/X_{Cr}^{melt} = 0.8 \pm 0.3$, where $X_{Cr}^{mnl}$ and $X_{Cr}^{melt}$ are wt% concentrations of Cr in minerals and silicate melt respectively (see above). Geophysical studies have shown that Mercury's crust is 20 to 50 km thick (e.g. Beuthe et al. 2020), which corresponds to 6% to 16% by mass of BSM. The abundance of Cr in BSM is calculated as a function of the average surface Cr found in this study ($X_{Cr}^{crust} = 200 \pm 60$ ppm):

$$X_{Cr}^{BSM} = X_{Cr}^{crust} * 0.06 + D_{Cr}^{mnl/melt} * X_{Cr}^{crust} * 0.94 \qquad (6)$$

Concerning the existence of sulfides in Mercury's interior, two scenarios have been suggested: the formation of an FeS melt during core formation due to the immiscibility between sulfides and metals (Malavergne et al., 2010) and the precipitation of Mg-Ca-rich sulfides from the crystallizing magma ocean or differentiation of the mantle (Malavergne et al. 2014, Boukaré et al., 2019). Therefore, here we considered three scenarios (Fig. 8): a sulfide-free Mercury, a model with FeS formed during core formation, and a model with Mg-Ca-rich sulfides formed during magma ocean crystallization. The core mass fraction was fixed to $f_{core} = 68\%$ (Hauck et al., 2013). For the second scenario, the mass fraction of possible FeS, $f_{sulf}$, was varied from 0 to 15% in substitution for the mantle, similarly to Boujibar et al. (2019). Mass fractions of 1, 5, 10 and 15% FeS would correspond to thicknesses of 14, 67, 131, and 191 km, respectively, during core-formation. These values would be overestimated if the sulfide layer is currently solid and has undergone significant compaction.

For the first two models (Fig. 8a-b), bulk Mercury (BM) Cr concentration was calculated by mass balance:

$$X_{Cr}^{BM} = X_{Cr}^{BSM} * \left(1 - f_{core} - f_{sulf}\right) + D_{Cr}^{\frac{met}{sil}} * X_{Cr}^{BSM} * f_{core} + D_{Cr}^{\frac{sulf}{sil}} * X_{Cr}^{BSM} * f_{sulf} \qquad (7)$$

We considered an equilibration between metal, silicate, and sulfide at the liquidus temperature of Mercury's mantle (2230 K) (Namur et al. 2016a). Other variables in the thermodynamic models (Eq. 3 & 5) such as the chemical composition of metal, silicate and sulfide were like those used in Boujibar et al. (2019). For the first model where sulfides are absent (Fig. 8a), these calculations were applied for a range of log $f$O$_2$ from IW-7 to IW-2. In the context of Mercury's core-mantle differentiation, sulfides are known to form as immiscible phases when the metal phase is enriched in Si (Morard & Katsura, 2010) or C (Corgne et al., 2008; Dasgupta et al., 2009). Indeed, the immiscibility field shrinks at higher pressure (> 15 GPa) while pressure at Mercury's core-mantle boundary (5.5 GPa) is low enough to allow for a large range of compositions where FeS- and FeSi- or FeC- rich liquids are immiscible. In addition, Mercury's bulk S reaches the upper estimates of S abundances in chondrites at IW–7, because of increased S solubility in magmas at low $f$O$_2$ (Namur et al. 2016a, Boujibar et al. 2019). The addition of sulfides at IW–7 would yield a sulfur abundance higher than in chondrites. Therefore, here we considered the possible presence of sulfides for models where the log $f$O$_2$ is between IW–6 and IW–2 (Fig. 8b).

For the third scenario which considers Ca-Mg-rich sulfides (Fig. 8c), we assumed that core formation happens as a first step in similar conditions as those described in the first model. We considered that following that step, as the magma ocean cools down, it equilibrates with the Ca-Mg-rich sulfides at a slightly lower temperature than during core formation (2000 K instead of 2230 K). Since S solubility in silicate melt decreases at lower temperature (Namur et al. 2016a), in the context of a reduced magma ocean with negligible Fe, exsolved sulfides are expected to be enriched in Ca and Mg. To model this scenario, after using Eq. 6 to calculate the Cr abundance in BSM, we calculated the Cr concentration in a Mercury magma ocean (MMO), and in bulk Mercury from:

$$X_{Cr}^{MMO} = (X_{Cr}^{BSM} * (f_{mant} + f_{crust}) + D_{Cr}^{\frac{sulf}{sil}} * X_{Cr}^{BSM} * f_{sulf})/(f_{mant} + f_{crust} + f_{sulf}) \quad (8)$$

$$X_{Cr}^{BM} = X_{Cr}^{MMO} * (1 - f_{core}) + D_{Cr}^{\frac{met}{sil}} * X_{Cr}^{MMO} * f_{core} \quad (9)$$

In this case, we consider that the magma ocean becomes sulfide-saturated as soon as it cools down right before its crystallization. If sulfide saturation happens at a later stage, only a fraction of the BSM would equilibrate with sulfides, and resulting bulk Cr/Al would be closer to the one calculated in the sulfide-free models (Fig. 9a). We considered a log $f$O$_2$ range from IW–6 to IW–4. The lower limit was based on the super-chondritic bulk sulfur (similarly to scenario 2, Boujibar et al. 2019) and the upper limit was fixed at IW–4 because of the low $f$O$_2$ required to permit the presence of stable Mg-Ca-rich sulfides (Namur et al., 2016a). We fixed the temperature at 2000 K (Boukaré et al., 2019) and we assumed that the sulfide phase contained 20 wt% Mg and 45 wt% S, i.e., the average composition of Mg-Ca-rich sulfides in Namur et al. (2016a). For all three scenarios, since there are significant uncertainties related to partition coefficients ($D_{Cr}^{mnl/melt}$, $D_{Cr}^{met/sil}$, and $D_{Cr}^{sulf/sil}$), surface Cr surface measurements ($X_{Cr}^{crust}$), and crustal thickness, we conducted Monte Carlo simulations to account for these errors. Random numbers were generated from normal distributions, which yielded $10^7$ models for Mercury differentiation for each of the considered combinations of $f$O$_2$ and sulfide mass fraction. We present in Fig. 8 the 68% most likely models (corresponding to one sigma standard deviation for a normal distribution) and discuss them in the following sections.

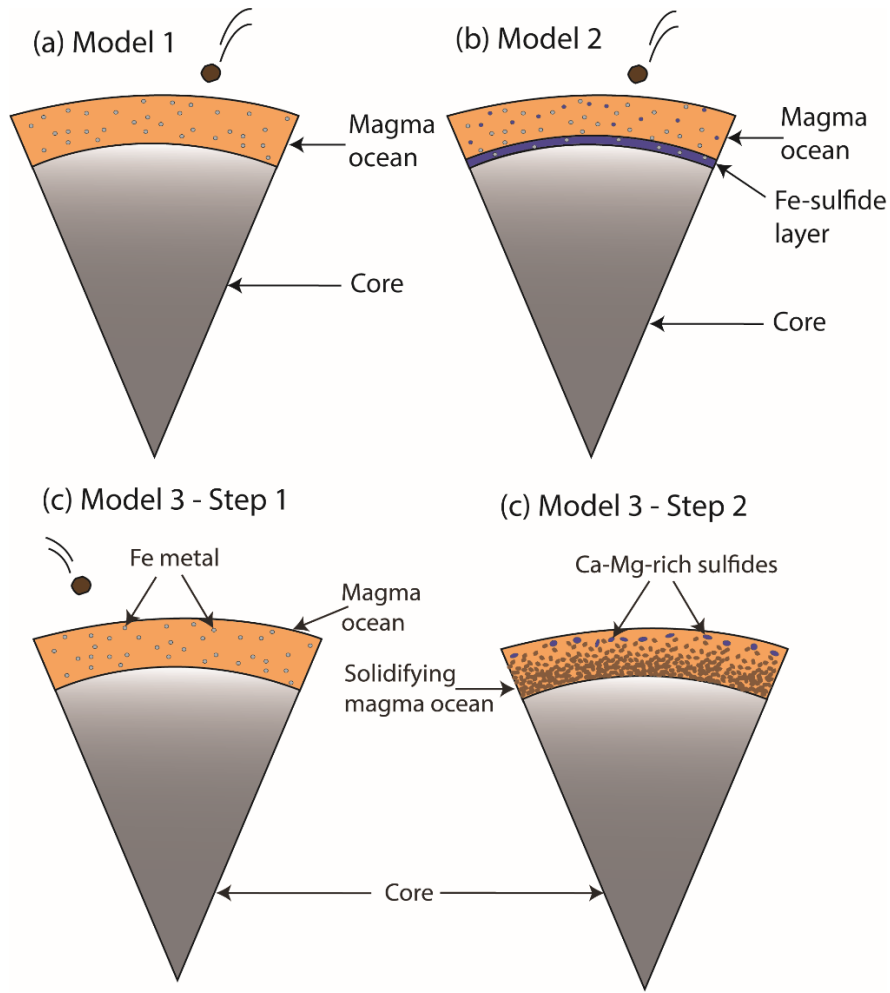

Fig. 8 (a) Model of Mercury's differentiation where sulfides are absent and droplets of Fe-rich metal fall in the magma ocean and equilibrate at the core-mantle boundary. (b) Differentiation model where iron sulfide phases are formed due to Mercury's enrichment in sulfur. Sulfides accumulate at the core-mantle boundary and form a FeS layer. (c) In the third model, core-mantle differentiation happens similarly to the model 1 (step 1). In this case however, as the magma ocean cools down, sulfur saturates in the magma ocean and Ca-Mg-rich sulfides form (step 2). The final distribution of sulfides in Mercury's mantle will depend on the density contrast between sulfide and silicate phases.

## 5. Implications for the oxygen fugacity and presence of sulfides during Mercury differentiation.

Our results show that the presence of Cr in Mercury's core explains the sub-chondritic Cr/Al ratio observed on the surface. Moreover, we find that the lower the $f$O$_2$, the higher the bulk Cr/Al that is computed for Mercury (Fig. 9)—i.e., because more Cr should be segregated into the core at lower $f$O$_2$ due to the increasingly siderophilic behavior of Cr in reduced conditions. We selected the 68% most likely values for bulk Mercury Cr/Al (equivalent to 1 sigma error for a normal distribution) for each combination of $f$O$_2$ and sulfide content in the Monte Carlo modeling. The results show that the bulk Mercury Cr/Al ratio matches chondritic values (0.0472 to 0.509) (Nittler et al., 2004) if the oxygen fugacity is between IW–6.5 and IW–2.5 in a sulfide-free system (Fig. 9a) and between IW–5.5 and IW–2 if Mercury has an FeS layer at its core-mantle boundary (Fig. 9b). If Mg-Ca-rich sulfides were present, the $f$O$_2$ range consistent with chondritic bulk Cr/Al would be narrowed to be between IW–5.5 and IW–4 (Fig. 9c) given the instability of Mg-Ca-rich sulfides at $f$O$_2$ above IW-4 (Namur et al., 2016a). If the $f$O$_2$ was close to the lower end and sulfides existed, they would only represent very small fractions of bulk Mercury (~1 wt% FeS or ~5 wt% Mg-Ca-S at IW–5), while with the highest $f$O$_2$ (IW–4), up to 15 wt% sulfide could have been present.

Other MESSENGER data have previously been used to constrain Mercury's oxygen fugacity. First, McCubbin et al. (2012) used the measured Mercury surface Fe abundance (Weider et al., 2014) and, by assuming that it is entirely present in the form of Fe$^{2+}$, these authors estimated the log $f$O$_2$ to be around IW–3 to IW–2.6. Later, by assuming that some Fe is present in a metallic form due to reaction with graphite, McCubbin et al. (2017) found lower values of log $f$O$_2$, ranging from IW–3.2 to IW–4.3. Namur et al. (2016a), however, suggested even lower $f$O$_2$ values, averaging IW–5.4. Their result was based on a comparison between the high S concentration of Mercury's surface (which is the most elevated among all terrestrial planets of our Solar System) and S solubility in magmas. In addition, Cartier et al. (2020) used MESSENGER XRS measurements of Ti on Mercury's surface, along with core formation models to show that Mercury could have a bulk chondritic Ti/Al ratio and the observed surface Ti/Al ratio if log $f$O$_2$ = IW–5.4 ± 0.4. However, the XRS-derived Ti abundance was not used to test whether other redox conditions would still reconcile surface compositions with chondrites. Boujibar et al. (2019) also showed that if Mercury's core was formed at $f$O$_2$ higher than IW–4, Mercury would not have enough Si in its core to yield a chondritic Fe/Si ratio, although this may be reconcilable for a CB-like Mercury bulk composition (Vander Kaaden et al., 2020). Finally, Anzures et al. (2020) suggested Mercury's mantle has log $f$O$_2$ between IW-4 to IW-2, based on the correlation of Ca and S concentrations observed at the surface of the planet and on the formation of CaS complexes in silicate melts at this range of $f$O$_2$ values. Our results, based on the surface Cr abundance, suggest a broad range of $f$O$_2$ conditions (IW–6 to IW–2), which overlap and are consistent with previously suggested ranges, including those based on surface Fe, S, and Si abundances.

Another important aspect of Mercury's differentiation is the possible existence of sulfides and their role in elemental fractionation. Figures 9b and 9c show the most likely bulk Cr/Al ratio (with associated errors) for different mass fractions (1, 5, 10, 15%) of the FeS layer (Fig. 9b) that may have precipitated during Mercury's core formation and of Mg-Ca-rich sulfides (Fig. 9c) that formed during magma ocean crystallization. Our results are similar whether sulfides are in the form of immiscible FeS or precipitated Mg-Ca-rich sulfides. The results also show that the presence of sulfides is possible if log $f$O$_2$ is higher than IW–5 ±0.5 (Fig. 9b). The lower the $f$O$_2$, the thinner the sulfide layer must be (or the smaller the amount of mantle sulfides must be) for bulk Cr/Al to be chondritic. Our results indicate that the sulfide layer may range in thickness from 67 km (if Mercury differentiated at IW–5) to 191 km (if the $f$O$_2$ was IW–4 or lower). Cartier et al. (2020) showed that at a log $f$O$_2$ of IW–5.4 ±0.4, the measured surface Ti/Al ratio would only be compatible with chondrites if sulfides were absent or at very low concentrations. We find a very similar result for Cr/Al at log $f$O$_2$=IW–5.5, but our results for higher $f$O$_2$ scenarios do not rule out the likelihood of the presence of sulfides in Mercury's interior.

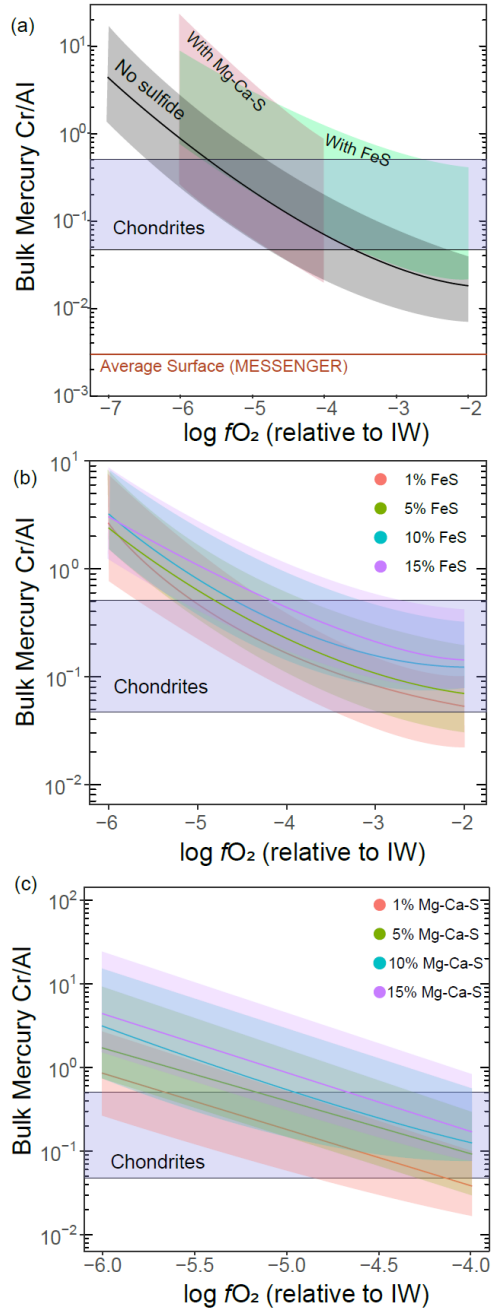

**Figure 9.** (a) Cr/Al ratios for chondrites, Mercury's surface, and three sets of models of bulk Mercury: with no sulfides (grey) and either the presence of a FeS layer at the core-mantle boundary (green) or Mg-Ca-rich sulfides in Mercury's mantle (yellow). The solid black line shows the most likely Cr/Al for a sulfide-free Mercury. The grey, green, and blue areas indicate the ranges of Cr/Al corresponding to 68% of the most likely results for each of the three scenarios. Models with different mass fractions, relative to bulk Mercury, of FeS and Mg-Ca-rich sulfides with associated errors are shown in (b) and (c), respectively.

## 5. Conclusion

We have reported the first systematic analysis of Cr abundances on Mercury's surface, based on data from MESSENGER's X-ray Spectrometer (XRS). The data indicate that, on average, Cr is present at a level of ~200±30 ppm (but with a possibly higher systematic uncertainty). The XRS data also indicate that Cr distributed heterogeneously across the planet and that it is correlated with major-element abundances. For example, the high-magnesium region (HMR) geochemical terrane has the highest observed Mg/Si, S/Si, Ca/Si, and Fe/Si ratios and also has a Cr/Si ratio 50% higher than Mercury's surface average. In contrast, the Caloris Basin (the largest recognized impact feature on Mercury) is depleted in all of these ratios relative to the planetary average (by 50% for Cr/Si). The average surface Cr/Al ratio is 0.003, some 17 to 170 times lower than that of chondritic meteorites.

We note that Vander Kaaden et al. (2017) used classical CIPW normative mineralogy calculations, modified to include the sulfides expected under highly reducing conditions, to constrain potential mineralogies consistent with measured surface elemental compositions across Mercury. Because of the lack of XRS data for Ti, Cr, and Mn available at the time that work was conducted, the authors performed two sets of calculations: one utilizing only reported values and the other assuming they were present at the XRS detection limits (e.g., 0.8 wt% for Ti, 0.5 wt% for Cr and Mn, Nittler et al., 2011) if abundance estimates were not available. The calculations that assumed the Ti, Cr, and Mn detection limits indicated that $TiS_2$, CrS, and MnS could make up some 50–70% of the sulfides present in the various regions. However, the significantly lower Ti (~0.2 wt%; Cartier et al., 2020) and Cr (200 ppm; this work) abundances determined since the work of Vander Kaaden et al. (2017) indicate that these calculations should be revisited.

We used a large set of published experimental data to explore the effects of temperature, pressure, oxygen fugacity, and composition on Cr partitioning between liquid silicates, sulfides, and metals. Combining these results with a planetary differentiation model, we found that to explain the sub-chondritic surface Cr/Al ratio with a bulk chondritic Cr abundance, Cr must be present in Mercury's core. Moreover, as seen previously from other oxybarometers, Mercury must have differentiated under highly reducing conditions. Our results indicate a broad range of redox conditions ($\log fO_2$ =IW–6.5 to IW–2.5) that are consistent with previous estimates based on surface Fe, S, and Si abundances. The presence of an FeS layer at the base of the mantle requires slightly less reduced conditions (log $f$O$_2$ =IW–5.5 to –2). The range is narrower — IW–5.5 to IW–4 — if substantial amounts of Mg-Ca-rich sulfides are present in the mantle. The existence of such sulfides, however, has been questioned based on surface Ti abundances that were derived from XRS data (Cartier et al., 2020). Additional abundance measurements for other elements on the surface, coupled with more precise experimental data and thermodynamic models are necessary to better estimate the oxygen fugacity of Mercury's interior. In the case of Cr, a better knowledge of its distribution between minerals and silicate melt at very low $f$O$_2$ would improve models of planetary differentiation. In addition, geochemistry instruments on the ESA/JAXA BepiColombo mission (e.g., Rothery et al., 2020), due to enter Mercury orbit in December 2025, are likely to provide Cr abundance estimates with broader spatial coverage, along with higher resolution and higher precision of Cr and thus will also provide better constraints on Mercury's oxygen fugacity.

## Acknowledgements

We thank the entire MESSENGER team for the development, launch, cruise, orbit insertion, and orbital operations of the MESSENGER spacecraft. We thank two anonymous referees for their constructive comments which helped improve this paper. This work was supported by the NASA Discovery Program under contract NAS5–97271 to The Johns Hopkins University Applied Physics Laboratory, NASW-00002 to the Carnegie Institution of Washington, NASA grant NNX07AR72G to LRN, and the Carnegie fellowship awarded to AB.

## Data Availability Statement

The MESSENGER X-ray Spectrometer data used in this work are available through NASA's Planetary Data System Geosciences node (https://pds-geosciences.wustl.edu/missions/messenger/xrs.htm; Starr 2018; Nittler 2018). Spectral fitting results used to generate maps in the paper as well as the maps themselves are available as Supplementary Information and are also deposited in Arizona State University's Research Data Repository (Nittler 2023). The experimental partitioning data are derived from the literature, and the complete list of references is given in the supplementary material. Custom IDL-based software was used to fit XRS spectra and generate elemental ratio maps. This software makes use of routines from the SolarSoft (Freeland & Handy, 1998) and MPFIT (Markwardt, 2009) libraries. The software is not compatible with the data products available through the PDS and is thus not publicly archived.

## Conflict of Interest Statement

The authors have no conflicts of interest to declare.

# Chromium on Mercury: New results from the MESSENGER X-Ray Spectrometer and implications for the innermost planet's geochemical evolution

Larry R. Nittler[1,2]*, Asmaa Boujibar[1,3], Ellen Crapster-Pregont[4,5], Elizabeth A. Frank[1], Timothy J. McCoy[6], Francis M. McCubbin[7], Richard D. Starr[8,9], Audrey Vorburger[4,10], Shoshana Z. Weider[1,11]

[1]Earth and Planets Laboratory, Carnegie Institution of Washington, Washington, DC, USA
[2]School of Earth and Space Exploration, Arizona State University, Tempe, AZ, USA
[3]Geology Department, Department of Physics & Astronomy, Western Washington University, Bellingham, WA, USA
[4]Department of Earth and Planetary Sciences, American Museum of Natural History, New York, NY, USA
[5]Department of Earth and Environmental Sciences, Columbia University, New York, NY, USA
[6]National Museum of Natural History, Smithsonian Institution, Washington, DC, USA,
[7]Astromaterials Research and Exploration Science Division, NASA Johnson Space Center, Houston, TX, USA
[8]Physics Department, The Catholic University of America, Washington, DC, USA
[9]Solar System Exploration Division, NASA Goddard Space Flight Center, Greenbelt, MD, USA,
[10]Physics Institute, University of Bern, Bern, Switzerland
[11]Agile Decision Services, Washington, DC, USA
*Corresponding author, lnittler@asu.edu.

**Contents of this file**

## Additional Supporting Information (Files uploaded separately)

**Maps.zip**  This is a zip archive containing  18 images (*.png) giving mapped Mg/Si, Al/Si and Cr/Si elemental abundance data on Mercury as shown graphically in Figure 3 as well as **Maps-README.txt**, a text file describing the *.png files.

**crflaredata.csv**: a table in comma-separated-variables format of fitting results for the 133 MESSENGER XRS flare data used in the paper. Cr/Si and Fe/Si ratios are corrected for the phase angle effect, as discussed in the paper. Both corrected and uncorrected data are provided for Cr/Si. Error bars are one-sigma. Errors are statistical and systematic error bars may be larger. Columns are defined as follows:

Flare: a number label for each record

STARTMET: Mission elapsed time (MET) of the first MXRS data record used for spectral sum. MET isseconds since launch of spacecraft.

ENDMET: Mission elapsed time (MET) of the last MXRS data record used for spectral sum. MET isseconds since launch of spacecraft.

START UTC: Date and time (in UTC) of the first MXRS data record used for spectral sum

END UTC: Date and time (in UTC) of the last MXRS data record used for spectral sum

PPLOTINCLUDE: If this variable is set to 1, this flare result was included in the calculation of phase angle correction and plotted in Figure 2

Area (km2): The total area of XRS footprint over spectral integration in square kilometers.

FPASPECT A unitless parameter that gives a rough measure of how asymmetric the XRS footprint is (high means stretched N-S, lower than 1 means stretched E-W); see L. R. Nittler et al., "Global major-element maps of Mercury from four years of MESSENGER X-Ray Spectrometer observations," Icarus, vol. 345, p. 113716, Jul. 2020, doi: 10.1016/j.icarus.2020.113716.

Latitude: average latitude of the XRS  footprint for the spectra integration.

Longitude: average latitude of the XRS  footprint for the spectra integration.

Incidence angle (deg): Average incidence (sun-surface normal) angle, in degrees, averaged over XRS footprint for spectral integration

Emission angle (deg): Average emission (XRS-surface normal) angle, in degrees, averaged over XRS footprint for spectral integration

Phase angle (deg): Average phase (spacecraft-surface-sun) angle, in degrees, averaged over XRS footprint for spectral integration

Solar temp (MK): Estimated average temperature of solar corona (in millions of kelvin) inferred from MESSENGER X-ray Solar Monitor measurement acquired at same time as spectral measurement

Mg/Si: Mg/Si weight ratio determined from fitting of spectral integration

er Mg/Si: One-sigma statistical error of Mg/Si weight ratio determined from fitting of spectral integration

Al/Si: Al/Si weight ratio determined from fitting of spectral integration

er Al/Si: One-sigma statistical error of Al/Si weight ratio determined from fitting of spectral integration

S/Si: S/Si weight ratio determined from fitting of spectral integration

er S/Si: One-sigma statistical error of S/Si weight ratio determined from fitting of spectral integration

Ca/Si: Ca/Si weight ratio determined from fitting of spectral integration

er Ca/Si: One-sigma statistical error of Ca/Si weight ratio determined from fitting of spectral integration

Ti/Si: Ti/Si weight ratio determined from fitting of spectral integration

er Al/Si: One-sigma statistical error of Ti/Si weight ratio determined from fitting of spectral integration

Cr/Si (not phase corrected): Cr/Si weight ratio determined from fitting of spectral integration (not phase corrected)

er Cr/Si (not phase corrected): One-sigma statistical error of Cr/Si weight ratio determined from fitting of spectral integration(not phase corrected)

Cr/Si (phase corrected): Cr/Si weight ratio determined from fitting of spectral integration (phase corrected)

er Cr/Si (phase corrected): One-sigma statistical error of Cr/Si weight ratio determined from fitting of spectral integration (not phase corrected)

Fe/Si: Fe/Si weight ratio determined from fitting of spectral integration (phase-corrected)

er Fe/Si: One-sigma statistical error of Fe/Si weight ratio determined from fitting of spectral integration (phase-corrected)

## Introduction

This document includes two supplementary figures referred to in the text and a table providing the literature references for the experimental partitioning data used in the manuscript (for example in Figures 6 and 7). Additional supplementary information included are:

### Text S1.

To construct a model predicting metal silicate partitioning of Cr, we performed a linear regression using 520 experimental data from 43 peer-reviewed publications (Ballhaus et al., 2013; Berthet et al., 2009; Boujibar et al., 2014, 2016, 2019, 2020; Cartier et al., 2020; Cartier, Hammouda, Boyet, et al., 2014; Cartier, Hammouda, Doucelance, et al., 2014; Chabot & Agee, 2003; Clesi et al., 2016; Corgne et al., 2008; C.R.M. Jackson et al., 2021; Dasgupta et al., 2013; Fischer et al., 2015, 2020; Geßmann & Rubie, 1998; Hiligren et al., 1994; Huang et al., 2020, 2021; Jana & Walker, 1997; Jennings et al., 2021; Kaaden & McCubbin, 2016; Kilburn & Wood, 1997; Laurenz et al., 2016; Malavergne et al., 2019; Mann et al., 2009, 2012; Righter et al., 2010, 2018; Siebert et al., 2011, 2012; Steenstra et al., 2017, 2018; Steenstra, Seegers, et al., 2020; Steenstra, Trautner, et al., 2020; Thibault & Walter, 1995; Tuff et al., 2011; Wade & Wood, 2001, 2005; Walker et al., 1993; Wood et al., 2008, 2014). Similarly, to model Cr partitioning between sulfide and silicate, we used a compilation of 253 experimental data from 17 peer-reviewed publications (Berthet et al., 2009; Boujibar et al., 2014, 2019, 2020; Cartier et al., 2020; Jana & Walker, 1997; Kaaden & McCubbin, 2016; Kiseeva & Wood, 2015; Laurenz et al., 2016; Mann et al., 2009; Namur et al., 2016; Steenstra et al., 2017, 2018; Steenstra, Haaster, et al., 2020; Walker et al., 1993; Wood et al., 2014). In the table S1 below we show these references in a table with additional information on C concentration used for modeling (see main text for more detail).

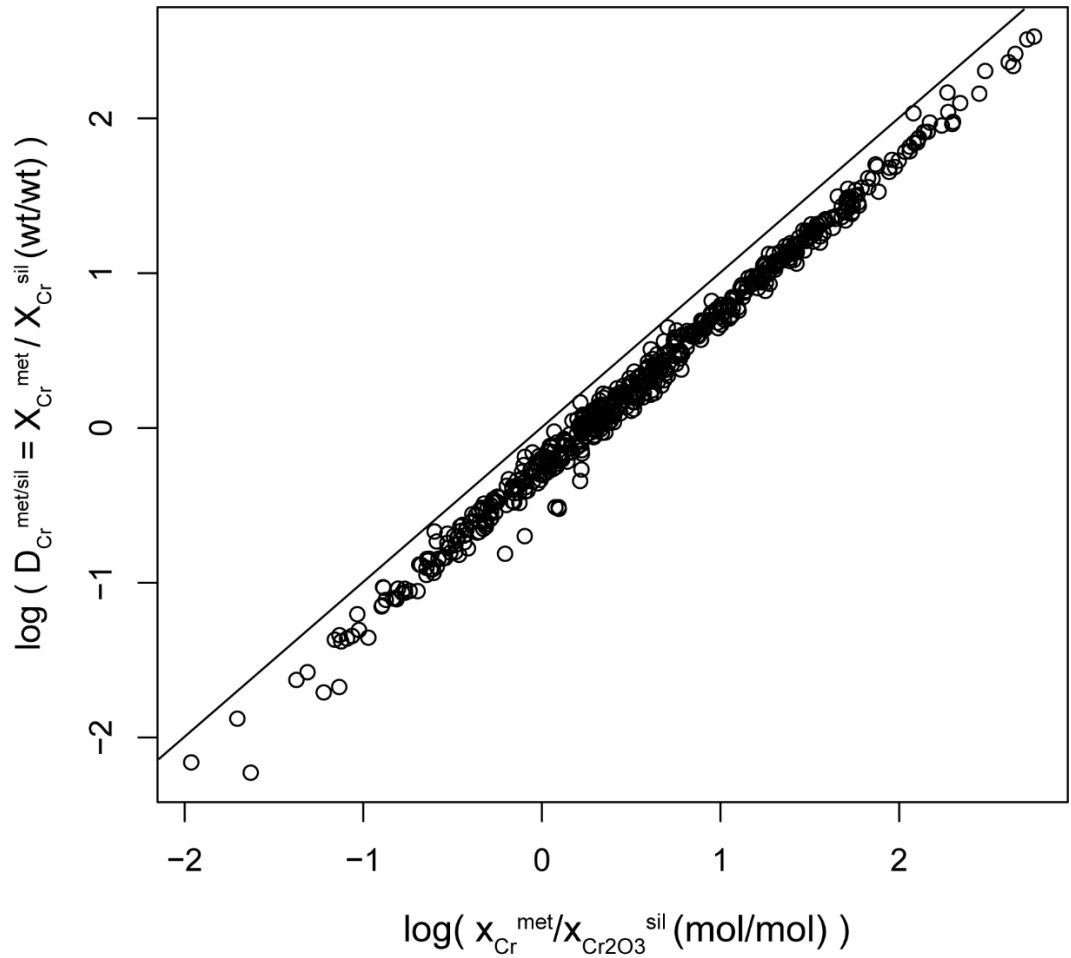

**Figure S1.** Comparison between the experimental Nernst partition coefficients $X_{Cr}^{met}/X_{Cr}^{sil}$ (wt/wt) with the molar $x_{Cr}^{met}/x_{Cr2O3}^{sil}$ partition coefficients. The black line represents y=x. The linear relationship between the log of both Cr partition coefficients calculated in these two different ways allows to model Nernst Cr partition coefficient using the equation (3) (see main text for more details).

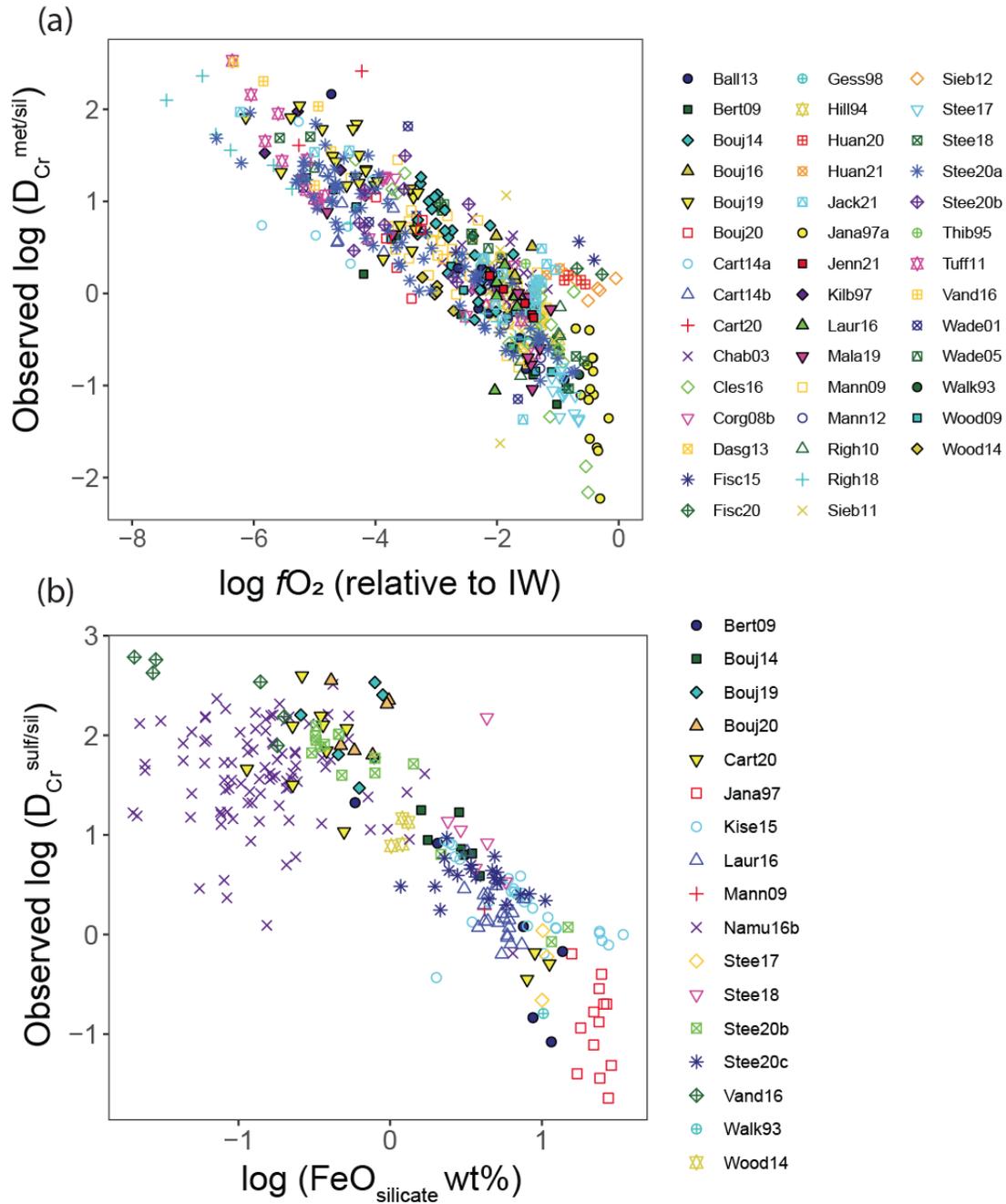

**Figure S2.** (a) Same as Figure 6b of Main Text, only with symbols indicating literature source of data points (Table S1). (b) Same as Figure 7b of Main Text, only with symbols indicating literature source of data points (Table S1).

**Table S1.** List of references for experimental data used in paper; codes are indicated on Figures 6, 7, and S2. A "*" following a reference indicates publications where experiments using graphite capsules have not measured C concentration in the metallic phase. In these experiments, C concentration was estimated by subtracting the sum of all other elemental concentrations from 100 wt%. The complete reference list can be found below at the end of this document.

| Metal-Silicate | | Sulfide-silicate | |
|---|---|---|---|
| Code | Reference | Code | Reference |
| Ball13 | Ballhaus et al. 2013 | Bert09 | Berthet et al. 2009* |
| Bert09 | Berthet et al.2009* | Bouj14 | Boujibar et al. 2014 |
| Bouj14 | Boujibar et al. 2014 | Bouj19 | Boujibar et al. 2019 |
| Bouj16 | Boujibar et al. 2016* | Bouj20 | Boujibar et al. 2020 |
| Bouj19 | Boujibar et al. 2019 | Cart20 | Cartier et al. 2020* |
| Bouj20 | Boujibar et al. 2020 | Jana97 | Jana & Walker 1997* |
| Cart14a | Cartier et al. 2014a* | Kise15 | Kiseeva & Wood 2015* |
| Cart14b | Cartier et al. 2014b* | Laur16 | Laurenz et al. 2016 |
| Cart20 | Cartier et al. 2020* | Mann09 | Mann et al. 2009* |
| Chab03 | Chabot & Agee 2003* | Namu16b | Namur et al. 2016* |
| Cles16 | Clesi et al. 2016* | Stee17 | Steenstra et al. 2017 |
| Corg08b | Corgne et al. 2008 | Stee18 | Steenstra et al. 2018* |
| Dasg13 | Dasgupta et al. 2013 | Stee20b | Steenstra et al. 2020c* |
| Fisc15 | Fischer et al. 2015 | Stee20c | Steenstra et al. 2020b* |
| Fisc20 | Fischer et al. 2020 | Vand16 | Vander Kaaden & McCubbin 2016* |
| Gess98 | Gessmann and Rubie 1998 | Walk93 | Walker et al. 1993* |
| Hill94 | Hillgren et al. 1994 | Wood14 | Wood et al. 2014 |
| Huan20 | Huang et al. 2020 | | |
| Huan21 | Huang et al. 2021 | | |
| Jack21 | Jackson et al. 2021 | | |
| Jana97a | Jana & Walker 1997* | | |
| Jenn21 | Jennings et al. 2021 | | |
| Kilb97 | Kilburn & Wood 1997 | | |
| Laur16 | Laurenz et al. 2016 | | |
| Mala19 | Malavergne et al. 2019* | | |
| Mann09 | Mann et al. 2009* | | |
| Mann12 | Mann et al. 2012 | | |
| Righ10 | Righter et al. 2010 | | |
| Righ18 | Righter et al. 2018 | | |
| Sieb11 | Siebert et al. 2011* | | |
| Sieb12 | Siebert et al. 2012 | | |
| Stee17 | Steenstra et al. 2017 | | |
| Stee18 | Steenstra et al. 2018* | | |

| Stee20a | Steenstra et al. 2020a* | | |
|---------|------------------------|---|---|
| Stee20b | Steenstra et al. 2020b* | | |
| Thib95 | Thibault & Walter 1995* | | |
| Tuff11 | Tuff et al. 2011 | | |
| Vand16 | Vander Kaaden & McCubbin 2016* | | |
| Wade01 | Wade & Wood 2001 | | |
| Wade05 | Wade & Wood 2005* | | |
| Walk93 | Walker et al. 1993* | | |
| Wood09 | Wood et al. 2009 | | |
| Wood14 | Wood et al. 2014 | | |